\documentclass[numberedappendix]{emulateapj}
\usepackage{natbib}
\slugcomment{Submitted to ApJ}

\shorttitle{Two Distinct Population in Dwarf Spheroidal Galaxies.}
\shortauthors{Kawata et al.}

\begin{document}


\title{Origin of Two Distinct Populations in Dwarf Spheroidal Galaxies}

\author{Daisuke Kawata\altaffilmark{1,2},
Nobuo Arimoto\altaffilmark{3}, Renyue Cen\altaffilmark{4},
and Brad K. Gibson\altaffilmark{2,5}
}

\altaffiltext{1}{The Observatories of the Carnegie Institution of Washington,
 813 Santa Barbara Street, Pasadena, CA 91101
\email{dkawata@ociw.edu}}
\altaffiltext{2}{
 Swinburne University of Technology, Hawthorn VIC 3122, Australia
}
\altaffiltext{3}{National Astronomical Observatory, 2-21-1 Osawa,
 Mitaka, Tokyo 181-8588}
\altaffiltext{4}{
Department of Astrophysical Sciences, Princeton University, Peyton
 Hall, Ivy Lane, Princeton, NJ 08544
}
\altaffiltext{5}{
Centre for Astrophysics, University of Central Lancashire, Preston,
Pr1 2HE, United Kingdom
}

\begin{abstract}
We study the chemical and kinematic properties
of the first galaxies that formed at high redshift,
using high resolution cosmological numerical simulations,
and compare them with the recent observational results 
for the Sculptor dwarf spheroidal galaxy of Tolstoy et al.,
who found two distinct stellar populations:
the lower metallicity stars are more spatially extended
and possess a higher velocity dispersion than the higher metallicity stars.
Our calculations reproduce these observations as the result of 
a steep metallicity gradient within a single population, 
induced by dissipative collapse of the gas component.
We also predict strong [N/O] enhancements in the lowest metallicity 
stars in dwarf spheroidals, due to the preferential retention of ejected gas
from intermediate-mass stars, compared to Type II supernovae.
\end{abstract}


\section{Introduction}
\label{sec-intro}

 Dwarf galaxies in the Local Group 
are one of the prime observational targets for galactic astronomy
because their relatively short distances enable
us to observe the properties of individual stars, which 
provide many details of their formation history
\citep{wb44}. Dwarf Spheroidals (dSphs) are one of the most populous
dwarf galaxy types seen in the Local Group and are
defined as galaxies with $M_B>-14$ mag, low surface brightness
($\mu_V>22$ mag arcsec$^{-2}$), no well-defined nucleus
(although in some dSphs, such as Fornax and Sagittarius, 
a globular cluster seems to correspond to a nucleus), 
and very little gas \citep{gw94,eg97,mm98}.
They do not have any ongoing star formation.
Some of them do not have any intermediate-age stars at all.
On the other hand, our current favorite $\Lambda$-dominated
cold dark matter ($\Lambda$CDM) cosmology suggests that
small objects form first, and larger systems are built up
by the assembly of smaller systems \citep[e.g.,][]{wr78}.
Therefore, it is considered that dSphs might be the first generation
of galaxies and survived cannibalization by larger systems.
Hence, dSphs might contain a record of the epoch of the
end of the dark age.
Recently, using a cosmological numerical simulation, \citet{rg05} 
demonstrated that
the small galaxies formed at high redshifts
can explain the global properties, such as the luminosity,
velocity dispersion, and iron abundance, of the dwarf galaxies 
observed in the Local Group. 
Their study encourages us to undertake further
investigation of the connection between the first galaxies and the dSphs.

With the wide-field multi-object 
spectrograph FLAMES on the Very Large Telescope,
\citet[][hereafter T04]{tihbj04} have measured the metallicity and
line-of-sight velocity for 300 member stars,
distributed over a large radial range ($\sim20$ times its core radius),
in the Sculptor (Scl) dSph galaxy.
They found that the stars in the Scl dSph show
two distinct populations. One of them is more metal 
rich ([Fe/H]$\sim-1.4$) with a centrally concentrated
distribution. The other one is metal-poor
([Fe/H]$\sim-2$) and more spatially extended.
In addition, the higher metallicity stars show lower velocity
dispersion than the lower metallicity stars.
This is unprecedented information for unveiling the formation
history of the Scl dSph.

To disentangle the formation process of the dSph from
such observational data, we construct
a theoretical model that can be compared with the
observations. 
 The aim of this paper is to show our first attempt to
make a self-consistent numerical simulation model that can 
be compared with the observational data of the Scl dSph 
directly and quantitatively.
We pay particular attention to the radial trend
of iron abundance and velocity dispersion, 
which are not discussed in \citet{rg05}, for comparison with
the unprecedentedly detailed observation presented in T04.
The next section describes our numerical method.
Section \ref{sec-res} presents our simulation results
and comparison with the observational data in T04.
Our discussion and conclusions are given in Section \ref{sec-disc}.

\section{Numerical Methods}
\label{sec-meth}

 In this study we focus on the properties of a dwarf galaxy
that formed at a high redshift in our high-resolution 
cosmological simulation.
The simulation was carried out using the galactic chemodynamics
code {\tt GCD+} \citep{kg03a}.
{\tt GCD+} is a three-dimensional tree $N$-body/smoothed
particle hydrodynamics (SPH) code that incorporates self-gravity,
hydrodynamics, radiative cooling, star formation, supernova (SN)
feedback, and metal enrichment. {\tt GCD+} takes account of chemical
enrichment by both Type~II (SNe~II) and Type~Ia (SNe~Ia) SNe and mass loss
from intermediate-mass stars, and follows the chemical enrichment history
of both the stellar and gas components of the system.
To study the formation process of small systems,
we update the code to implement 
non-equilibrium chemical reactions of hydrogen and
helium species (H, H$^{+}$, He, He$^{+}$, He$^{++}$, H$_{2}$,
H$_{2}^{+}$, H$^{-}$) and their cooling processes.
The details of the updated code are described in the Appendix \ref{sec-code}. 
%Here, we briefly describe our cosmological simulation models. 

We adopt a $\Lambda$CDM cosmology of $\Omega_0 h^2 = 0.135 $, 
$\Lambda_0=1-\Omega_0$, $\Omega_{\rm b} h^2 = 0.0224$, and $h=0.71$ 
\citep{svpkn03}, and use a
multi-resolution technique \citep{kg03b}
to achieve high-resolution in the regions of
interest, while outer regions exerting the tidal forces 
are handled with lower resolution.
The initial conditions for the simulations are constructed
using the public software {\tt GRAFIC2} \citep{eb01}. 
Gas dynamics and star formation are included only within 
the relevant high-resolution region ($\sim$80~kpc in comoving scale); 
the surrounding low-resolution region
($\sim$530~kpc diameter sphere) 
contributes to the high-resolution region only through gravity. 
 Consequently, the initial condition consists of a total of 287,491
dark matter particles and 233,280 gas particles.
The mass and softening length of individual gas and dark matter
particles in the high-resolution region are $129$
$650$ ${\rm M}_\odot$ and 30 and 51 pc, respectively.

\begin{figure}
%\epsscale{.80}
\plotone{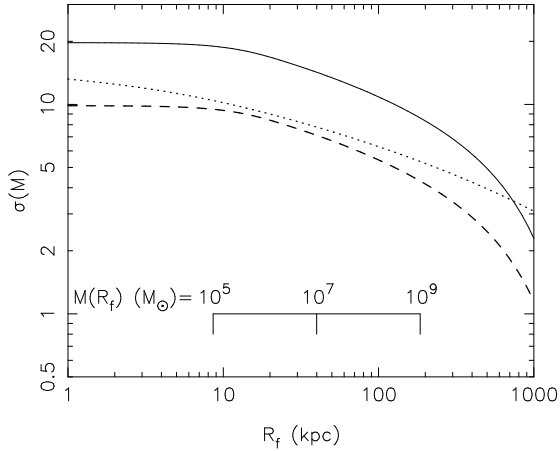}
\caption{
The mass variance for the CDM power spectrum as a function
of filter radius, R$_{\rm f}$. The mass corresponding to the
filter radius, M(R$_{\rm f}$), is also shown in the panel
The dotted line shows the mass variance with $\sigma_8 = 0.9$.
The dashed line indicates the same as dotted line, but
the power spectrum for the wave-length longer than 
our simulation volume ($\sim$530 kpc) and shorter than Nyquist
wave-length ($\sim$22 kpc) are ignored.
The solid line demonstrates the same as the dashed line,
but with $\sigma_8=1.8$, i.e.\ our applied power spectrum.
\label{fig-powers}}
\end{figure}

 To reduce the computational cost, we applied a significantly
small simulation volume. Hence, our simulation misses the
density perturbation induced by larger scale modes.
Figure \ref{fig-powers} demonstrates that the mass variance
$\sigma(M)$ becomes slightly smaller at the mass scale
in which we are interested ($\sim10^8 M_{\rm \odot}$)
if the power spectrum for wavelengths longer than our 
simulation volume are ignored.
In addition, the first galaxies are expected to
form at high-density peaks, i.e.\ biased regions \citep{mw96}.
Therefore, we adopt a higher value of $\sigma_8=1.8$,
instead of the value suggested by recent observations ($\sigma_8=0.9$).
Figure \ref{fig-powers} also shows the mass variance 
for the applied power spectrum.
The adopted high $\sigma_8$-value enables us to form
a relevant dSph-like small galaxy in the particular realization
with the relatively small simulation volume, in addition to partly
compensating for the missing waves larger than our simulation box.

\begin{figure*}
%\epsscale{.80}
\plotone{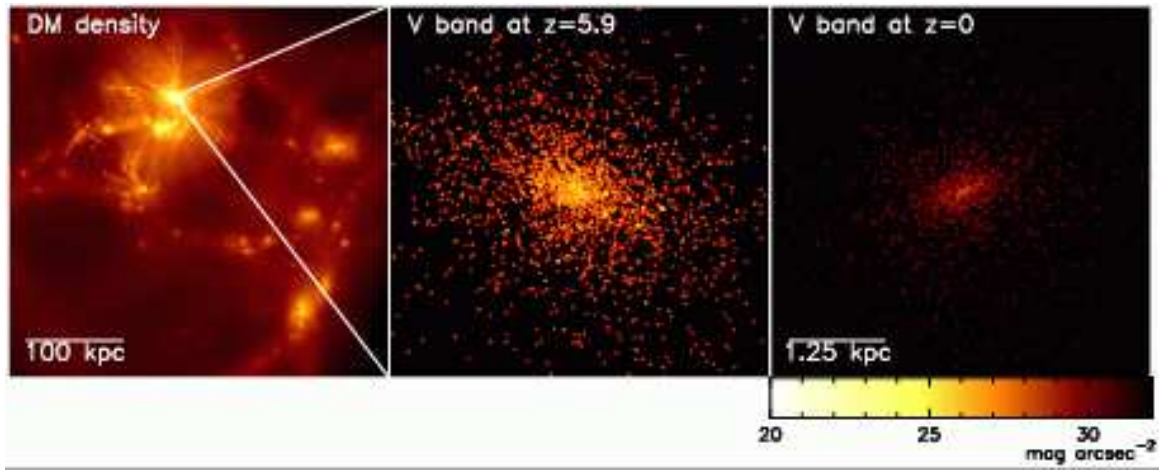}
\caption{
 Dark matter density distribution at $z=5.9$ (left), 
$V$-band (rest-frame) luminosity distribution at $z=5.9$, and 
expected $V$-band luminosity distribution at $z=0$ after the passive evolution.
\label{fig-dmv}}
\end{figure*}

 In the high-resolution region, we find a small stellar system
at $z=5.9$ (Figure \ref{fig-dmv}). 
The virial radius and mass of this system 
are respectively 1.9 kpc and $5.1\times10^7 {\rm M}_\odot$ at $z=5.9$. 
Here, we follow the fitting formula in the Appendix of \citet{ks96}
to define the virial mass and radius, taking into account
the cosmology and redshift.
We assume that at this redshift, star formation in
this system has been quenched by mechanisms, such as
re-ionization \citep[e.g.,][]{ge92,cn94,tw96,bkw00,bflbc02,su04a}
 and/or galactic wind \citep[e.g.,][]{ds86,ay87},
and that the system evolves
passively afterwards. Thus, we assume that the chemical and kinematic
properties at $z=5.9$ would not change until $z=0$, and we analyze
the properties expected at $z=0$ from the output of the simulation
at $z=5.9$. 
This is, of course, consistent with our assumption that the observed
dSph is a galaxy that formed early and retained its own identity until z=0. 
We check this {\it Ansatz} by examining its detailed properties.

\begin{figure}
%\epsscale{.80}
\plotone{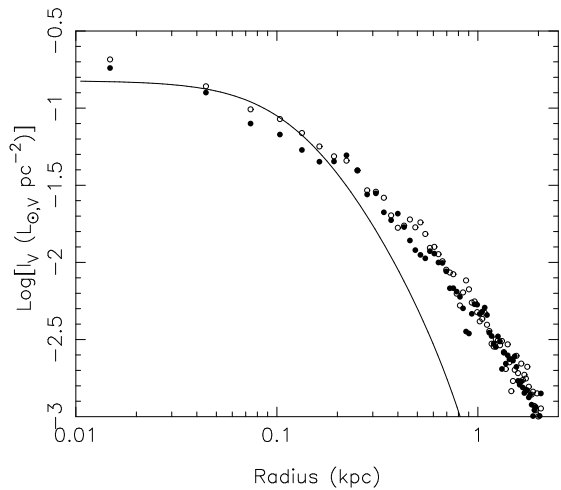}
\caption{
 The $V$-band surface brightness profile from the simulation data
before (open circles) and after (solid circles) the relaxing run 
(see text for details). The solid line presents a King profile
with the core radius, $r_c=0.12$ kpc, and tidal radius, $r_t=1.6$ kpc.
\label{fig-sfb}}
\end{figure}

The middle panel of Figure \ref{fig-dmv} shows
the rest-frame $V$-band luminosity distribution of this galaxy at $z=5.9$.
The luminosity distribution is calculated by our 
population synthesis code,
%\citep[which
%itself is based upon the population synthesis models of][]{ka97},
taking account of the metallicity and age of the star particles
at $z=5.9$. We use the single stellar population spectrum data 
in \citet{ka97}. The data do not include
any emission lines. For simplicity, we do not take into account any absorption
by the inter-stellar (ISM) and inter-galactic medium (IGM).
To make sure that the system is dynamically stable and
that the distribution of stellar population and kinematics
does not change dramatically until $z=0$,
we run a pure N-body simulation using the particles within 
a radius of $\sim1.7$ times the virial radius at $z=5.9$,
as an initial condition. We run this N-body
simulation (relaxing run) for about 5 Gyr (corresponding to the time
from $z=5.9$ to $z=1$).
% and confirmed that the system
%is indeed dynamically stable, and the properties
% analyzed in this paper are almost unchanged.
The right panel of Figure \ref{fig-dmv} shows the rest-frame $V$-band
luminosity distribution analyzed from the final output of 
this relaxing run extrapolated at $z=0$, based on the stellar age at $z=0$.
Figure \ref{fig-sfb} shows the $V$-band surface brightness
profiles for the system before ({\it open circles}) and 
after ({\it filled circles}) the relaxing run.
The profile after the relaxing run is similar to
the profile before. We also confirm that all the properties
presented in this paper are not changed after the relaxing run.
Therefore, the properties at $z=5.9$ are expected to be unchanged
till $z=0$. We present only the results from the data
after the relaxing run in the next section.

The above relaxing run does not explicitly include the effects of
two-body relaxation, because the system's stellar density is so low that
its effects are not dominant.  We have derived the total number of stars
within the virial radius, $N_{\rm s}$, the velocity dispersion,
$\sigma_{0}$, and the number density of stars, $n_{\rm s,0}$, taking into
account the initial mass function (IMF) and the remnant stars, in the
central region, and obtained $N_{\rm s}\sim6.5\times10^5$, $\sigma_0\sim10$
km s$^{-1}$ and 
$n_{\rm s,0}\sim0.005$ pc$^{-3}$ with a mean stellar mass of about
$m_*=0.3$ M$_{\sun}$. The two-body relaxation time can then be estimated as
\citep[e.g.,][]{ko65}
\begin{equation}
 t_r = \sigma_0^3/(4\pi G^2 m_*^2 n_{\rm s,0} ln N_{\rm s}).
\end{equation}
This leads to $t_r = 7.0\times10^{14}$ yr, suggesting that two-body 
relaxation is not important for this system.

 The solid line in Figure \ref{fig-sfb} represents a King profile
with the same core radius $r_{\rm c}$
and tidal radius $r_{\rm t}$ as observed in the Scl dSph,
i.e.\ $r_{\rm c} = 0.12$ kpc and $r_{\rm t}=1.6$ kpc.
We assume that the distance of the Scl dSph is 72 kpc \citep[][T04]{kd77}.
The central surface brightness is normalized to roughly match
the simulated surface brightness profile.
In the inner region, the simulated system has
a profile similar to the observed profile of the Scl dSph,
which means that the simulated galaxy has a similar core radius.
In the outer region, the simulated galaxy has a higher surface
brightness compared to the King profile. Therefore, the tidal
radius of the simulated galaxy is inconsistent with 
the observed one. However, our simulated galaxy is an isolated system,
while the Scl dSph is a satellite galaxy of the Milky Way.
We expect that if the simulated galaxy falls into a bigger host galaxy,
the stars in the outer region are tidally stripped, and thus that
the tidal radius would depend on their environment, which
could make the tidal radius of the simulated galaxy smaller and
similar to the observed one. 
Here, we assume that such stripping does not change the properties of 
the inner region, and compare the properties of this isolated system
with those of the Scl dSph.

 The basic properties of the simulated galaxy are summarized in 
Table \ref{tab-bmr}. In the next section, we compare
our model results with the observational data of
the Scl dSph in T04.

\begin{figure*}
%\epsscale{.80}
\plotone{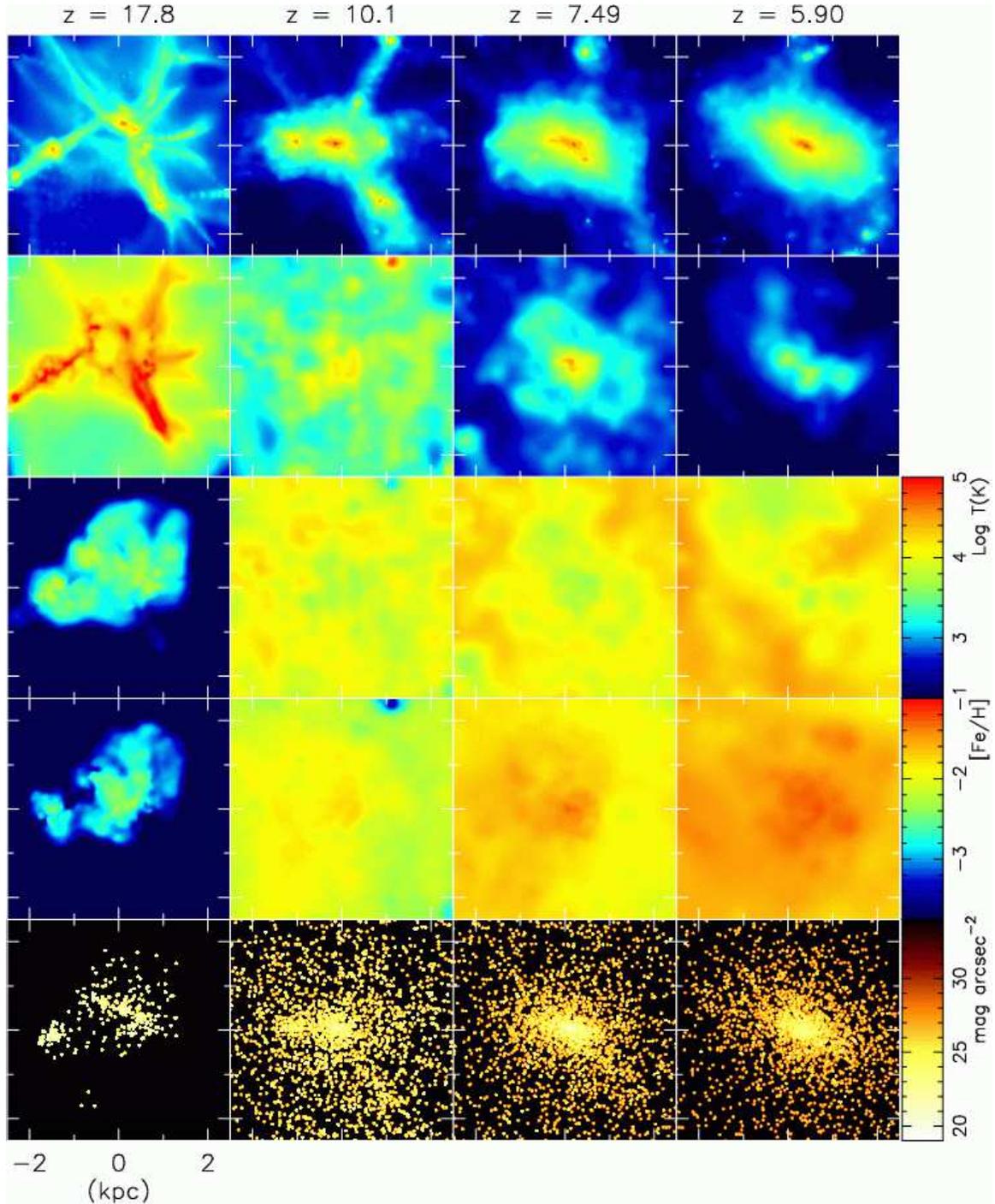}
\caption{
 Evolution of the distributions of dark matter density (top),
the gas density (2nd), the gas temperature (3rd), 
the iron abundance of gas (4th), and $K$-band (observed-frame)
luminosity (bottom).
\label{fig-evol}}
\end{figure*}

\begin{figure}
%\epsscale{.80}
\plotone{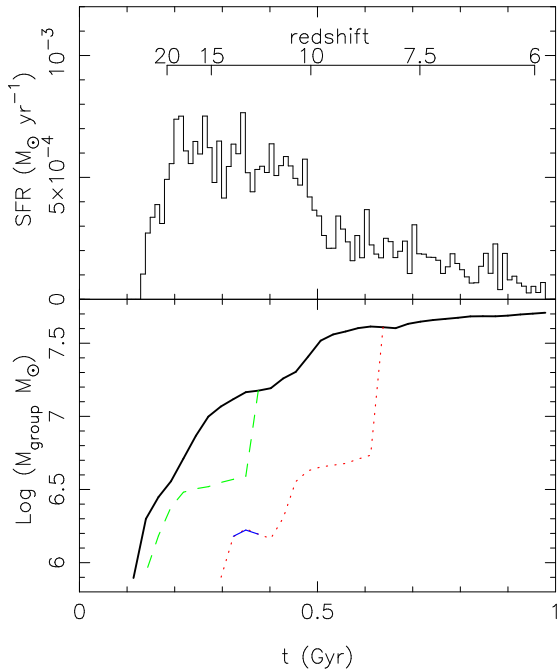}
\caption{
 Time variation of the star-formation rate and the evolution of
the virial mass of the progenitor halos. 
In the bottom panel, different styles of the lines
(black thick-solid, red dotted, green dashed and blue thin-solid)
indicate different halos. The connection between lines
displays that two halos merge together. For example,
the halo shown by red dotted line merges into the large
halo described by black thick-solid line at $t\sim0.64$ Gyr.
\label{fig-sfr}}
\end{figure}

\begin{figure*}
\plotone{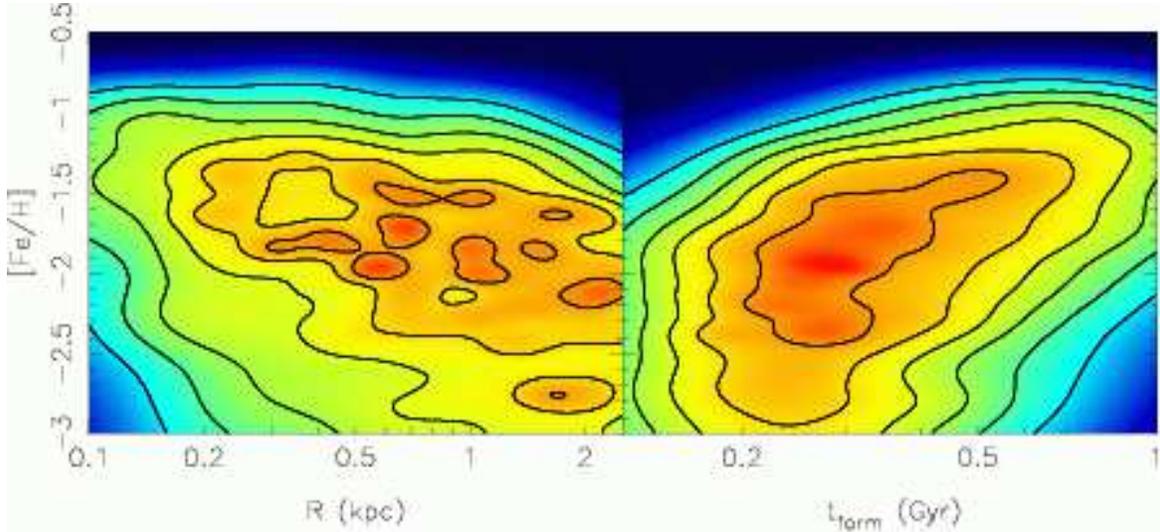}
\caption{
 Smoothed stellar mass distribution in the [Fe/H] vs.\ radius plane 
(left) and in the [Fe/H] vs.\ formation time, t$_{\rm form}$,
 plane, i.e. the age-metallicity relation (right).
\label{fig-fehr}}
\end{figure*}

\begin{figure}
%\epsscale{.80}
\plotone{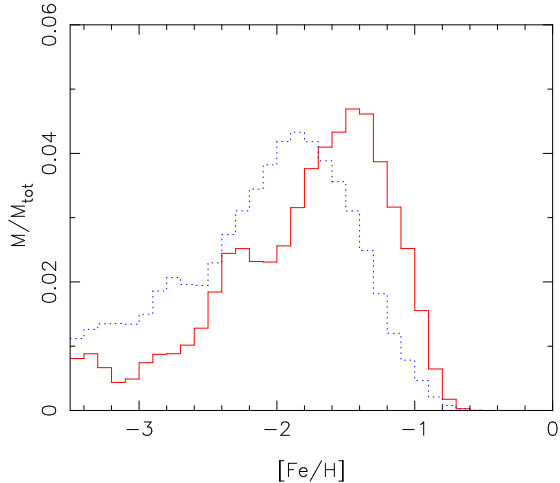}
\caption{
 Metallicity distribution function of stars 
in the inner ($R<0.25$ kpc: red solid histogram) and 
outer ($R>0.25$ kpc: blue dotted histogram) region.
\label{fig-mdf}}
\end{figure}

\begin{figure}
%\epsscale{.80}
\plotone{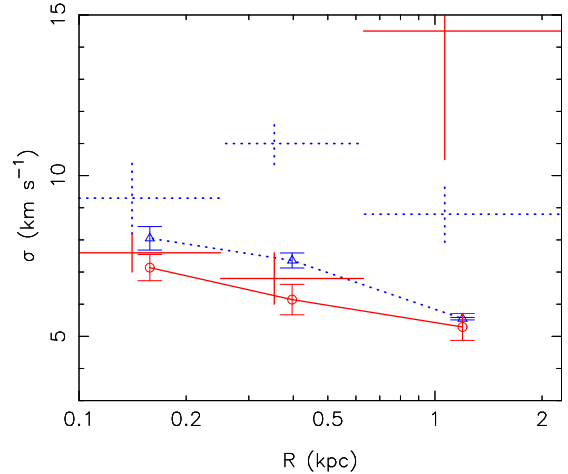}
\caption{
 Velocity dispersion profile of high ([Fe/H]$>-1.7$: 
red circles connected with solid line) 
and low ([Fe/H]$<-1.7$: blue triangles connected with
dotted line) metallicity stars. The error-bars represent
the dispersions from the measurements at the different projections.
The observational data in T04 are also shown as the red solid 
(for [Fe/H]$>-1.7$) and blue dotted (for [Fe/H]$<-1.7$) crosses, where
the horizontal-bar corresponds to the radial range, and
the vertical-bar corresponds to their dispersion.
The vertical-bars are plotted with an offset, for clarity.
\label{fig-sigr}}
\end{figure}

%\begin{figure}
%%\epsscale{.80}
%\plotone{fsfr.ps}
%\caption{
% The history of star formation rate normalized by the total mass
%for stars in the inner ($R<0.25$ kpc: solid histogram) and 
%outer ($R>0.25$ kpc: dotted histogram) region.
%\label{fig-fsfr}}
%\end{figure}

\begin{figure}
%\epsscale{.80}
\plotone{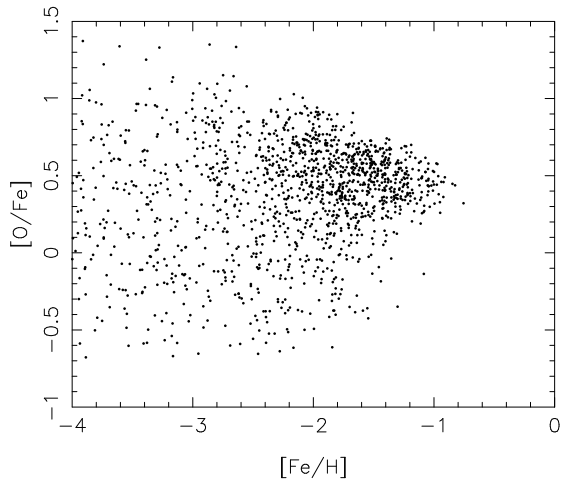}
\caption{
 [O/Fe] as a function of [Fe/H] for star particles within $R=2.5$ kpc.
\label{fig-ofe}}
\end{figure}

\begin{figure*}
\plotone{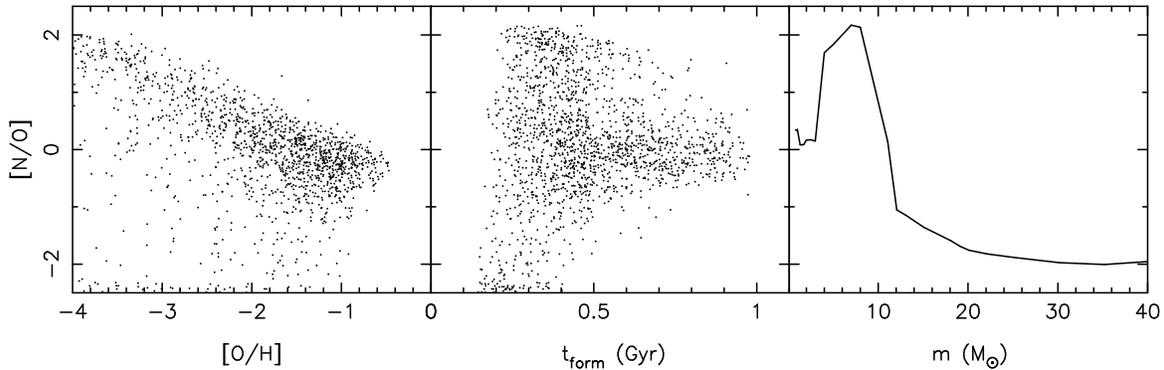}
\caption{
 {\it Left}: [N/O] as a function of [O/H] for star particles within
 $R=2.5$ kpc. {\it Middle}: [N/O] as a function of their formation
epoch. {\it Right}: [N/O] yields as a function of the mass
of the progenitor stars with metallicity of $Z/Z_{\sun}=0.01$
\citep{ww95,vdhg97}.
\label{fig-no}}
\end{figure*}

%%%%% Table 1

\begin{deluxetable*}{rrrrrrr}
\tablecolumns{6}
\tablewidth{0pc}
\tablecaption{Basic Model Results \label{tab-bmr}}
\tablehead{
 \colhead{M$_{\rm vir}$} & \colhead{r$_{\rm vir}$} &
 \colhead{M$_{\rm gas}$($<{\rm r}_{\rm vir}$)} & 
 \colhead{M$_{\rm DM}$($<{\rm r}_{\rm vir}$)} & 
 \colhead{M$_{\rm star}$($<{\rm r}_{\rm vir}$)} & 
 \colhead{M$_{V}$} &  \colhead{T$_{\rm vir}$\tablenotemark{a}}  \\
 \colhead{($M_{\sun}$)} & \colhead{(kpc)} &
 \colhead{($M_{\sun}$)} & \colhead{($M_{\sun}$)} & \colhead{($M_{\sun}$)} &
 \colhead{(mag)} & \colhead{(K)} 
}
\startdata
$5.1\times10^7$ & 1.9 &
$3.5\times10^5$ & $5.0\times10^7$ & $1.9\times10^5$ &
$-7.23$ & $2.8\times10^3$
\enddata
\tablenotetext{a}{Virial temperature calculated by $G{\rm M}_{\rm vir}\mu m_p/3
 k_{\rm B} r_{\rm vir}$ \citep{ks96}.}
\end{deluxetable*}

%%%%% Table 2

\begin{deluxetable}{rrrrrrrrr}
\tablecolumns{9}
\tablewidth{0pc}
\tablecaption{ Heavy element mass budget for the system within r$_{\rm vir}$.
 \label{tab-mej}}
\tablehead{
 \colhead{} & \multicolumn{4}{c}{$z=12.1$} &  \colhead{} & 
 \multicolumn{3}{c}{$z=5.9$} \\ \cline{2-5} \cline{7-9} \\
 \colhead{} & 
 \colhead{${\rm M}_{Z,\rm ej}$\tablenotemark{a}} & 
 \colhead{${\rm M}_{Z,\rm g}$\tablenotemark{b}} & 
 \colhead{${\rm M}_{Z,\rm s}$\tablenotemark{c}} & 
\colhead{${\rm f_{\rm esc}}$\tablenotemark{d}} & \colhead{} & 
 \colhead{${\rm M}_{Z,\rm ej}$} &
 \colhead{${\rm M}_{Z,\rm s}$} & 
\colhead{${\rm f_{\rm esc}}$\tablenotemark{e}} \\
 \colhead{} & \colhead{($M_{\sun}$)} & \colhead{($M_{\sun}$)} &
 \colhead{($M_{\sun}$)} & \colhead{} & \colhead{} & 
  \colhead{($M_{\sun}$)} & \colhead{($M_{\sun}$)} &
 \colhead{}
}
\startdata
N & 17.0  & 7.6 & 0.26 & 0.55 & &
 160.0  & 7.2 & 0.96 \\
O & 170.0 & 57.0 & 1.3 & 0.66 & &
 1900.0  & 74.0 & 0.96 \\
Fe & 17.0 & 2.3 & 0.066 & 0.86 & &
 150.0 & 3.0 & 0.98 \\
\enddata
\tablenotetext{a}{Total ejected mass.}
\tablenotetext{b}{Mass in gas component.}
\tablenotetext{c}{Mass in stellar component.}
\tablenotetext{d}{Escape fraction, 
$1-({\rm M}_{Z,\rm g}+{\rm M}_{Z,\rm s})/{\rm M}_{Z,\rm ej}$.}
\tablenotetext{e}{Escape fraction, 
$1-{\rm M}_{Z,\rm s}/{\rm M}_{Z,\rm ej}$.}
\end{deluxetable}

\section{Results}
\label{sec-res}

 Figure \ref{fig-evol} shows the formation
history of this galaxy. Although some minor mergers
are involved, the system forms through smooth accretion.
%The stars are forming in the region close to the center.
The lower panel of Figure \ref{fig-sfr} shows the virial mass evolution
of the galaxy and building blocks that merge into the galaxy.
The panel demonstrates that the main system experiences only two minor mergers
(mass ratio smaller than 0.4). 
The top panel of Figure \ref{fig-sfr} shows 
the history of the star formation rate (SFR).
In this galaxy, star formation starts at $z=25.5$
and the SFR has a peak around $z=15$. 
The $z=17.8$ panel in Figure \ref{fig-evol} shows that
the gas is blown out and the heavy elements are distributed to 
the IGM.
Table \ref{tab-mej} presents the heavy-element mass
budget for the stars that are within the virial radius 
at $z=5.9$. The total ejected mass M$_{Z, \rm ej}$
means the mass of each heavy element ejected from stars within r$_{\rm vir}$
until that redshift. The mass in the stellar component, M$_{Z, \rm s}$,
is defined as the mass of each heavy element still within the stellar component
within the virial radius. We define the escape fraction 
as ${\rm f}_{\rm esc}=1.0-{\rm M}_{Z, \rm s}/{\rm M}_{Z, \rm ej}$. 
Here we assume that the heavy elements in the gas component
are also blown out after some mechanism stops star formation at $z=5.9$.
Table \ref{tab-mej} indicates that more than 96\% of the heavy
elements produced in stars have escaped from the system until $z=5.9$.
Our assumed SNe feedback ($7.5\times10^{50}$ erg per supernova,
which is chosen to reproduce the low metallicity of the Scl dSph)
has a strong effect on the gas dynamics
and continuously blows out the gas from the system.
However, the continuous gas accretion leads to further
star formation, albeit at a somewhat lower rate.
Figure \ref{fig-sfr} shows that star formation continuously
occurs even around this blow-out phase.
Nevertheless, the star formation of this small system
is strongly suppressed by SNe feedback, which helps
to keep the stellar metallicity low, as seen in the age-metallicity
relation shown in the right panel of Figure \ref{fig-fehr}.

 As described in Section \ref{sec-intro}, T04
found that the stars in the inner region have 
higher metallicity than those in the outer region.
They split the two regions at a radius of about 0.25 kpc
and demonstrated that the metallicity distribution function (MDF)
of stars in the inner (outer) region has a peak around [Fe/H]$=-1.4$
($-2.0$).
To compare with their result, Figure \ref{fig-mdf} shows 
the MDF for stars in the inner ($R<$0.25 kpc\footnote{
Throughout this paper, 
$R$ represents radius at an arbitrary projection, and
we ignore the stars $R>2.5$ kpc.})
and outer (R$>$0.25 kpc) regions for our simulated galaxy.
The MDF for the inner (outer) region of the simulated galaxy
has a peak at [Fe/H]$\sim-1.4$ ([Fe/H]$\sim-1.9$),
which is in good agreement with the observed MDFs in T04.
Therefore, the simulated galaxy also show two distinct
stellar populations.

We found that this is simply due to the metallicity
gradient in the system.
Figure \ref{fig-fehr} shows that the peak of the metallicity distribution 
at different radii gradually decreases as the radius increases,
i.e., that there is a metallicity gradient in the simulated system. 
Since the metallicity gradient is steep enough in the system,
the MDF for the inner region moves toward higher [Fe/H],
compared to the MDF for the outer region.
This demonstrates that the steep metallicity gradient 
can cause two different chemical properties
in the inner and outer regions.

Next, we analyze the kinematic properties. Figure \ref{fig-sigr}
presents the velocity
dispersion at different radii for the low-metallicity ([Fe/H]$<-1.7$)
and high-metallicity
([Fe/H]$>-1.7$) stars. This velocity dispersion is obtained by
taking the line-of-sight velocity dispersion of stars within annuli at
different radii. To improve the statistics, we analyzed the line-of-sight
profiles at 36 different projections, and obtained the mean values and
dispersions, which are represented as error-bars in the 
figure. 
Figure \ref{fig-sigr} also shows the observed velocity dispersion of
the low-metallicity ([Fe/H]$<-1.7$) and high-metallicity
([Fe/H]$>-1.7$) stars in T04.
Within a radius of about 0.6 kpc, 
the observational data show that the velocity dispersion
of the high-metallicity stars is lower than that of the
low-metallicity stars.\footnote{Note that T04 mentioned that
in the region outside of the radius of 0.6 kpc, the number
of the high metallicity stars are too small.}
In other words, the two different metallicity components
also have distinctive dynamical properties.
Our simulation results also show the same trend. Although the difference
is small, the difference is significant and more than twice that of the
dispersions shown as the error bars in the inner region.

Our simulation demonstrates that
a system formed at a high redshift can reproduce
the two stellar populations whose chemical and dynamical properties
are distinctive.
 However, the simulated galaxy shows some inconsistent results
with the observed properties of the Scl dSph.
First, compared with Figure 3 of T04, the MDFs
for both inner and outer regions of the simulated galaxy
have too long a tail at lower [Fe/H]. 
In the observational data, there
are no stars at [Fe/H]$<-2.8$, although T04 selected their samples from 
the limited region of the color-magnitude diagram, which might
tend to exclude stars of too low and too high metallicity.
On the other hand, the simulated galaxy has a significant 
fraction of stars with such low metallicity.
Next, Figure \ref{fig-sigr} also shows that the velocity dispersion
of our simulated galaxy is too small compared with
the observed values. In addition, Table \ref{tab-bmr} shows that the
$V$-band magnitude of the simulated galaxy ($M_V=-7.23$)
is also small, compared with the luminosity
of the Scl dSph ($M_V=-10.7$). 
These are problems of our current model that need to be
solved in a future study, and we discuss possible solutions
in the next section.

 Finally, we have also analyzed abundance ratios.
\citet{et05} shows that stars in the Scl dSph 
show different distributions in the [$\alpha$/Fe] versus
[Fe/H] plane, compared with the stars in the solar neighborhood
\citep[see also][]{svtph03,gswgc05}.
In the Scl dSph, the [$\alpha$/Fe] values of member stars with 
[Fe/H]$<-2$ are higher than solar abundance ratios, and 
[$\alpha$/Fe] values approach the solar value as [Fe/H] 
increases at [Fe/H]$>-2$.
On the other hand, in the solar neighborhood [$\alpha$/Fe] values are 
constantly higher than the solar value for stars with [Fe/H]$<-1$ 
and starts decreasing at [Fe/H]$=-1$.
This difference can be explained by 
the contribution from SNe Ia, which decreases [$\alpha$/Fe], and 
a lower star formation rate in the dSph, which keeps [Fe/H] low
until the chemical enrichment by SNe Ia becomes important
\citep[e.g.,][]{ia02,lm03,lm04,lmc05}. However, Figure \ref{fig-ofe} 
demonstrates
that in the simulated galaxy, the mean [O/Fe] is almost constant
(oxygen is one of the typical $\alpha$-elements).
One reason why there is no enrichment from SNe Ia
is because we implemented an SNe Ia model proposed 
by \citet{ktn00}, who suggested
that SNe Ia are inhibited in stars with [Fe/H]$<-1$.
Even if we relax their [Fe/H] limit for SNe Ia, it is 
difficult to explain the observational trend.
\citet{ktn00} consider that the mass ranges of companion
stars for the SNe Ia progenitor binaries are restricted
to between 0.9 and 1.5 M$_{\rm \sun}$ (they call this the RS+WD system) and
to between 1.8 and 2.6 M$_{\rm \sun}$ (the MS+WD system).
The expected lifetime of 2.6 M$_{\rm \sun}$ stars with 
Log($Z/Z_{\rm \sun}$)$=-2.3$ are about 0.5 Gyr,
according to the lifetime used in \citet{ka97}.
On the other hand, Figures \ref{fig-sfr} and \ref{fig-fehr}
indicate that stars with [Fe/H]$>-2$ start forming even
0.1 Gyr after the first star formation.
However, the progenitors of SNe Ia are still unknown.
For example, another SNe Ia progenitor model suggested by 
\citet{gr83} suggests that the mass range of binaries is
between 8 and 3 M$_{\rm \sun}$. Since the lifetime of such stars
is 0.03-0.3 Gyr, if we applied their model,
the observational trend of [O/Fe] and [Fe/H] could be explained.
Another solution to explain the decrease in [$\alpha$/Fe]
is increasing the contribution of SNe II from stars with masses
between 10 and 15 M$_{\sun}$ whose yields provide
[$\alpha$/Fe]$<0$ \citep[e.g.,][]{ww95,bg97}.
However, the observed trend of the decreasing [$\alpha$/Fe] 
requires a fine-tuned variation of the shape of the IMF,
depending on metallicity, which seems unlikely. 

 Figure \ref{fig-ofe} also shows that the simulated [O/Fe]
has too large a scatter, compared with the observational data.
The particles with a low [Fe/H] are likely
enriched only a few times, and then their 
abundance pattern reflects that of the yields from 
the stars within a small mass range.
As a result, the scatter of [$\alpha$/Fe] for the star particles 
becomes as large as the [$\alpha$/Fe] variation
of the yields of SNe II progenitors with different masses \citep{ww95}. 
This indicates a serious problem of the current chemical evolution model
in particle-based numerical simulations.
The problem is also known in explaining the small dispersion
of [$\alpha$/Fe] for metal-poor halo stars 
([Fe/H]$=-3$ to $-2$) in the Milky Way, which
provides a strong constraint to the chemical evolution history 
of the Galaxy \cite[e.g.,][]{aranb05}.
So far, numerical simulations of disk galaxy formation
show clearly larger scatter than what is observed \citep[see also][]{rvn96}.
This suggests that we have to consider some metal-mixing model
between particles. Nevertheless, the mean values of metallicity from 
the current simulation models may be relatively robust.

 We also analyzed the abundance ratio of [N/O] against [O/H]
and found an interesting feature.
Figure \ref{fig-no} demonstrates that there is a visible trend
that [N/O] decreases from [N/O]$=2$ to $\sim0$ as [O/H] increases
from [O/H]$=-4$ to $-1.5$. The middle panel of the figure
also shows that high [N/O] are seen in the stars formed around $t=0.3$ Gyr
and that stars born at later epochs have lower [N/O] values.
This high [N/O] comes from stars with masses between 4 and
8 M$_{\sun}$ whose lifetime is around 0.1 Gyr 
(Fig.\ \ref{fig-no}, {\it right}).
To make the contribution from such intermediate-mass stars important,
we are required to suppress the contribution from higher mass stars
whose [N/O] values are lower and whose lifetimes are shorter. 
As seen in Figure \ref{fig-evol},
the enriched gas is blown out at a high redshift around $z=18$,
due to the strong feedback by SNe II and the relatively shallow
potential of the system at such a high redshift.
As a result, chemical enrichment by massive stars,
i.e.\ SNe II, becomes less important and
enrichment from intermediate-mass stars becomes relatively more important.
As the system becomes larger and the gravitational potential binds the
gas enriched by SNe II, [N/O] starts decreasing.
%This is also demonstrated in Table \ref{tab-mej} which shows 
%the metal budget at $z=12.1$ and $z=5.9$ (see above for details). 
This is also demonstrated in Table \ref{tab-mej}, which shows 
the metal budget at $z=12.1$ and $z=5.9$. Note that
since star formation is still in progress,
the escape fraction at $z=12.1$ is defined as 
${\rm f}_{\rm esc}=
1.0-({\rm M}_{Z, \rm g}+{\rm M}_{Z, \rm s})/{\rm M}_{Z, \rm ej}$.
Here, M$_{Z, \rm g}$ is the mass of each heavy element 
still within the gas component within the virial radius. 
% after proof
At $z=12.1$, the escape fraction of oxygen
is significantly higher than that of nitrogen, and 
the values become closer at $z=5.9$.
Oxygen is mainly produced in SNe II, although it is still created
in the intermediate-mass stars \citep{vdhg97}.  
On the other hand,
% after proof
nitrogen mainly comes from the intermediate-mass stars.
Thus, Table \ref{tab-mej} indicates that the gas enriched by SNe II
is preferentially blown out at high redshift. 
Therefore, testing this trend of [N/O] in dSphs would be interesting.
To our knowledge, the nitrogen abundance has not been observed
in dSphs. On the other hand, the intermediate-mass stars are also
expected to be progenitors of $s$-process elements.
Recent observational studies 
\citep[e.g.,][]{sm02,tvsph03,sai04,ms05a,ms05b} suggest that the
$s$-process elements are enhanced in the dSphs. 
This might be explained by
a process in dSph in which the SNe II-enriched gas is 
preferentially blown out of the system 
and the enrichment by intermediate-mass stars
is more important.
Interestingly, \citet{sm02,ms05b} have shown that
the $s$-process elements are more enhanced in higher metallicity stars
in the Sagittarius dSph.
In addition, \citet{sai04} found a very metal-poor star
([Fe/H]$=-2.7$) that has an anomalously low abundance of $s$-process
elements in the Ursa Minor dSph.
This may indicate a delay of $s$-process element enrichment,
because $s$-process elements come from relatively low-mass 
(1.5-3 M$_{\rm \sun}$), i.e. long-lifetime, stars
 \citep[see also][]{lmc05}.

\section{Discussion and Conclusions} 
\label{sec-disc}

 We have analyzed chemical and kinematic properties of 
a small system that formed at high redshift in a $\Lambda$CDM
cosmological simulation. Our simulated galaxy shows that higher metallicity
([Fe/H]$>-1.7$) stars have a more centrally concentrated distribution and 
lower velocity dispersion, compared with the lower metallicity stars
([Fe/H]$<-1.7$). This trend is consistent with 
the observed trend in the Scl dSph reported in T04. 
Thus, we conclude that 
a survivor of a small system that formed at high redshift
can explain the observed stellar chemical and kinematic properties.

 T04 claim that this observed trend indicates that
there are two distinct populations in the Scl dSph.
They propose three possible mechanisms to explain the two populations. 
(1) Two episodes of star formation.
Subsequent SNe feedback from the initial star formation
blows out the gas and stops star formation temporally.
After SNe feedback becomes weaker,
more metal-rich gas comes back and forms a new generation of 
stars \citep{ccgl01,mfm02}.
(2) External influences, such as minor mergers or accretion of
additional gas at a later epoch. (3) Selected heating by UV background
radiation. The radiation evaporates the outer layers of the gas
preferentially, and star formation lasts longer in the inner
region \citep{su04b}. 
Figure \ref{fig-sfr} shows that our simulated galaxy does not
have two episodes of star formation. 
In this study, we assume that
there is no later minor merger or gas accretion after star formation
is stopped at high redshift, and star formation is abruptly terminated
without any time delay depending on the radius.
Thus, the formation history of our simulated galaxy 
does not correspond to any of the above scenarios. 

Figures \ref{fig-evol} and \ref{fig-sfr} show that the simulated galaxy 
forms through smooth accretion rather than major mergers. 
It is known that such smooth accretion of the dissipative gas component
leads to the higher metallicity for the gas in the inner region, 
because gas that dissipatively accretes into the inner region 
is enriched by stars in the outer region, as well as by 
stars in the inner region, where the stellar density is higher.
As Figure \ref{fig-sfr} shows, the substantial duration of the starburst at 
$z=13\sim20$ allows for significant self-enrichment in
the inner region of the galaxy, resulting in a higher metallicity
stellar population there. The outer regions are progressively
less vigorous in star formation activities due to the supernova explosion
feedback effect, which significantly reduces the gas content of
the galaxy, as well as subsequent merger subunits. The latter point
is clearly visible in Figure \ref{fig-sfr};
for example, the merger event at $z\sim8$ is
not associated with any increased star formation activity, indicating
a lack of gas.
This mechanism makes the mean metallicity of the stars in the inner
region higher. 
This is the same mechanism that is suggested to
explain the metallicity gradient for larger spheroidals, i.e.\
normal elliptical galaxies \citep{rl74,rc84,kg03a,ck04}. 
In fact, Figure \ref{fig-fehr} shows 
a metallicity gradient in the simulated system. 
Hence, our simulation demonstrates that 
for a small system that stopped forming stars at a high redshift,
it is possible to have a metallicity gradient, which can
explain the {\it apparent} two distinct populations
observed in the dSph Scl stars. Note that 
the metallicity gradient has to be steep, to make such a difference
in the inner and outer regions. Thus, if the steepness of
the metallicity gradient is responsible for the apparent
two populations in the dSph and metallicity gradients
are different between dSphs, as seen between elliptical
galaxies \citep[e.g.,][]{ka99}, 
there may be dSphs that do not show two such populations,
due to a less steep metallicity gradient. Thus, it would be
important to obtain more samples of both simulated and
observed dSphs. 
However, here we also note that our simulation result 
does not rule out alternative scenarios to explain the observed two
populations, such as those discussed in T04.
 
 Unfortunately, some properties
of our simulated galaxy are inconsistent with those observed
in the Scl dSph.
Figure \ref{fig-mdf} shows that in both the inner and outer regions
there are significant amounts of stars that have extremely
low metallicity. 
%The stars formed around $z=20$ 
%seen in Figure \ref{fig-fsfr} correspond to those extremely
%metal-poor stars which have not been observed in the Scl dSph.
This is the same problem as the so-called G dwarf problem,
known from the difficulty 
for a simple chemical evolution model 
to explain the MDF of the G dwarfs in the solar neighborhood 
\citep[e.g.,][]{svb62,bt75} and to reproduce
the spectrum features in the central 
region of bright elliptical galaxies \citep[e.g.,][]{ay87,bcf94,lg97}.
The solutions suggested for the solar neighborhood G dwarf problem
include (1) gas infall, (2) prompt initial enrichment (PIE), and
(3) metal-enhanced star formation (MESF). The infall model
has so far been very successful in solving the local G dwarf
problem, as it is likely that the disk of the Milky Way has been formed
by continuous accretion of gas from reservoirs, such as 
Galactic halo gas and the IGM. However, our simulation
already takes into account cosmological gas infall. Hence,
the infall model likely does not work for interpretation of 
the MDF of the dSph.
However, we also note that our simulation does not take into account the 
effect of ionizing radiation fields.  If there is a background field, 
regardless of its specific origin, some fraction of the molecular 
hydrogen will be photodissociated \citep{on99,mba01,rgs01,yahs03}.
As such, our simulation likely overestimates  H$_2$ 
cooling and, thus, the gas accretion rate.  
%Future enhancements to {\tt GCD+} will include a 
%phenomenological background ionizing radiation field, an expressed 
%purpose for which will be to explore its role in ameliorating the G-dwarf 
%problem.

The MESF model is also unlikely, as it is also fully taken into
account in our simulations by introducing the radiative cooling rate
depending on the metallicity of the gas particles. This leaves the PIE
scenario as the most attractive possibility for solving the G dwarf
problem of dSph galaxies. 
The possible mechanism would be that 
at high redshift ($z\sim20$), Population III stars formed 
at the center of the building blocks and supernova
explosions blew up the gas in the building blocks,
which helped to enrich the IGM.
Then the enriched gas fell back into the system, which became
larger after multiple mergers of building blocks,
and Population II stars were produced from the enriched gas. 
If the lifetime of the lowest mass Population III stars  
is shorter than a Hubble time, Population III
stars do not exist in the current dSph.
Although the first building blocks have to be
large enough ($\sim10^{5}$-$10^{6}$ $M_{\rm \sun}$) 
to make stars \citep[e.g.,][]{tsr97,aanz98,fc00,yahs03},
one massive Population III star is enough to blow out the gas
\citep[e.g.,][]{byh03}. Also, note that the explosion does
not have to enrich the IGM of the entire universe; it only needs to
enrich the nearby IGM, which falls back into the system.
If this is the case, the oldest stars in the dSph have
the signature, i.e.\ abundance pattern, of the first 
supernova explosion \citep[e.g.,][]{bkpfg03}. Hence, low-metallicity
stars in dSphs may be the best target
for looking for the signature of the first stars \citep[e.g.,][]{sai04}.

%A simple solution to solve this problem is ``hiding''
%these stars. If the lower limit of the mass for these 
%extremely low metallicity stars is as low as the mass of stars
%whose lifetime is shorter than the Hubble time, they will disappear at $z=0$.
%In this case, their mass should not be too high, and should not
%produce too many SNe, because they might 
%blow out the gas from the system completely \citep{byh03}.

 Another problem in our simulated galaxy is 
that its luminosity and velocity dispersion are too low.
We have tried different parameter sets of models of star formation
and SNe feedback.
However, we found that to keep metallicity as low as what is observed,
strong SNe feedback that leads to 
a low efficiency of star formation is required, 
which makes it difficult to produce enough stars at high redshift.
The simplest solution is that tens of such small systems
merge without any additional star formation. 
However, this is unlikely, and such mergers make
the metallicity gradient shallower \citep{sw80}.
Another solution would be to make the galaxy more massive.
The virial mass of our galaxy is $5\times10^7$ M$_{\rm \sun}$
at $z\sim6$. Dynamical analysis of dSphs suggests
that their mass-to-light ratio is more than 100
\citep[e.g.,][]{kweg01,hntsq03}, which 
suggests that the total mass of the Scl dSph is more than
$10^8$ M$_{\rm \sun}$. If the system were big enough, it might
be able to continue a low level of star formation even
after the re-ionization in the inner region,
along scenario 3 described above.
If this were the case, it would help
to solve ``the missing satellite problem'' \citep{kkvp99,mgg99}
as discussed in \citet{hntsq03,mcs05a}.
Admittedly, we have been restricted to the analysis of a single 
simulation, derived from one initial condition.  While this simulation 
does demonstrate one possible mechanism for explaining the origin of the 
two distinct populations in the Sculptor dwarf spheroidal, it is critical 
that the next step in this work entail a suite of 
simulations at comparable resolution,
each with different initial conditions and different 
patterns of small-scale perturbations.  This will allow the analysis of 
a suite of simulated dwarfs of differing masses and differing assembly 
histories.  Such a suite will allow our preliminary conclusions to be 
placed on firmer statistical ground, and will allow us to provide further 
insights into the link between mass, assembly history, and present-day 
observable characteristics.

 Our simulation demonstrates that the recently 
available detailed properties of the observed dSphs provide
invaluable information about their formation history.
Our current cosmological simulations obviously miss,
or oversimplify, some important physics.
For example, our SNe feedback model is too
simple, and/or the resolution of our simulation is not good enough
to describe SNe feedback, although the present study shows that
the effect of SNe would be crucial to explain
the observed properties of dSphs. 
Also, the IMF might depend on the physical condition of the progenitor gas,
such as metallicity. 
In addition, our simulation ignores radiative transfer effects,
such as radiative heating and pressure.
Strong light from new born stars would suppress cooling of the surrounding
gas, which would be important in a small system 
\citep{rgs02a,rgs02b,ky05}.
Nevertheless, dSphs are good laboratory for studying the
physical process of the galaxy formation. We hope that the present
study will be a good starting point for testing formation scenarios
for dSphs by comparing detailed observations
with chemo-dynamical galaxy formation models.

\acknowledgments

We thank Andrew McWilliam for his helpful advice during
the completion of this manuscript and the anonymous referee for
constructive comments.
This work is partly supported by NSF AST 02-06299, AST 04-07176, 
NASA NAG5-13381,
and a Grant-in-Aid for Scientific Research
(16540223) from the Japanese Ministry of Education, Culture, Sports, Science,
and Technology.
The financial support of the JSPS, through a
postdoctoral fellowship for research abroad, and
the Australian Research Council, through
its Discovery Project and Linkage International schemes, 
is gratefully acknowledged.
We acknowledge 
the Astronomical Data Analysis Center of the National Astronomical
Observatory, Japan (project ID wna15b), the
Institute of Space and Astronautical Science 
of the Japan Aerospace Exploration Agency, and
the Australian and Victorian Partnerships for Advanced
Computing, where the numerical computations for this paper were
performed.
\appendix

\section{Updated Version of GCD+}
\label{sec-code}

 This study focuses on a small system forming at 
high redshift, using a high-resolution cosmological simulation. 
To follow the physics on such small scale,
we have updated our original galactic chemodynamical evolution code,
{\tt GCD+}. In addition, we have changed some parameter values in
the code from our previous studies \citep[e.g.,][]{kg03a,kg03b}.
Below we explain what has been updated.

\subsection{Radiative cooling with 
non-equilibrium chemical reaction of hydrogen and helium species}

In the updated version,
the code follows non-equilibrium chemical reactions of hydrogen and
helium species (H, H$^{+}$, He, He$^{+}$, He$^{++}$, H$_{2}$,
H$_{2}^{+}$, H$^{-}$) and their cooling
processes. We assume ionization equilibrium cooling for
radiative cooling of heavy elements, because it is too expensive
to follow non-equilibrium chemical reactions of 
all the ionization states of elements heavier than helium
\citep{ko00}.
Thus, the total cooling rate is the summation of
the equilibrium cooling of heavy elements and
non-equilibrium cooling of hydrogen and helium.
The equilibrium cooling of heavy elements is calculated
using a code based on the Raymond-Smith code \citep{rs77},
which is used in \citet{ckor95}.

 We follow the non-equilibrium chemical reactions of 
species related to hydrogen and helium, i.e.\
H, H$^{+}$, He, He$^{+}$, He$^{++}$, H$_{2}$, H$_{2}^{+}$,
H$^{-}$, and e$^{-}$, and calculate their radiative cooling rate,
based on the method of \citet{aazn97,azan97}.
Although \citet{aazn97} used the molecular hydrogen cooling rate
of \citet{ls83}, we adopt an updated one suggested by \citet{gp98}.
The time-scales of radiative cooling and
chemical reaction of these species are much 
smaller than the dynamical time-scale
\citep{azan97,yahs03}.
Therefore, we use sub-time steps to calculate the chemical reaction
and integrate the thermal equation. 
The code considers the time-scale of the chemical reaction
to be represented by the time-scale of the change in 
number of free electrons,
$\tau_{\rm sp} = n_{\rm e}/\dot{n}_{\rm e}$. 
The time-scale for radiative cooling is calculated
by $\tau_{\rm rad} = e/|\Lambda-\Gamma|$, where $e$ is thermal
energy, and $\Lambda$ and $\Gamma$ are
the cooling and heating\footnote{In this paper, we do not
take into account any background radiation. Thus, there is
no photo-ionization heating.} rate, respectively.
 For each gas particle $i$, we set the sub-time step to be
$\Delta t_{{\rm sub},i} = \tau_{\rm dyn}/2^n\leq
\min(\epsilon \tau_{\rm sp},\epsilon \tau_{\rm rad})$
with $\epsilon = 0.1$, for efficient integration.
Here $\tau_{\rm dyn}$ is the minimum time step 
required from dynamical evolution,
$\Delta t_{\rm dyn}$ \citep{dk99}. 
Following \citet{azan97}, we update
the number $n_{k,i}$ of the $k$-th species in a gas particle $i$
by a backward differentiation method,
$n_{k,i}^{t+\Delta t_{{\rm sub},i}}=(C_{k,i}^{t+\Delta t_{{\rm sub},i}}
\Delta t_{{\rm sub},i}+n^t_{k,i})/(1+D_{k,i}^{t+\Delta t_{{\rm sub},i}}
\Delta t_{{\rm sub},i})$. Here, $C_{k,i}$ and $D_{k,i}$ are the creation 
and destruction rates, respectively. The thermal energy is updated
by a semi-implicit method in our code. Note that \citet{azan97} use 
the explicit method.  
However, we find that the semi-implicit method
is more robust (K. Yoshikawa 2003, private communication). 
We solve the following equation with iteration:
\begin{equation}
 u_i^{t+\Delta t_{{\rm sub},i}}  =  u_i^{t}  
 +\Delta t_{{\rm sub},i} 
 \frac{\Delta E_{{\rm SN},i}(t_n,\Delta t_{\rm dyn})}{\Delta t_{\rm dyn}}
 + 0.5 \Delta t_{{\rm sub},i}
%\\ \nonumber
\left[
 \left(\frac{du}{dt}\right)_{{\rm ad},i}^{t_n} 
+  
 \left(\frac{du}{dt}\right)_{{\rm ad},i}^{t_n+\Delta t_{\rm dyn}}
+ \frac{\Gamma_i^{t} -\Lambda_i^{t}}{\rho_{{\rm g},i}}
%
%\right.
%\\ \nonumber
%
 +
 \frac{\Gamma_i^{t+\Delta t_{{\rm sub},i}}
 -\Lambda_i^{t+\Delta t_{{\rm sub},i}}}{\rho_{{\rm g},i}}\right].
\end{equation}
Here, $\Delta E_{{\rm SN},i}(t_n,\Delta t_{\rm dyn})$ is 
the heating energy from SNe within the dynamical time step
from $t_n$ which indicates the time at the beginning of 
this sub-time step integration.
Also, $(du/dt)_{{\rm ad},i}$ is the adiabatic term \citep{dk99} for 
the thermal equation.
Since $(du/dt)_{{\rm ad},i}$ requires a neighbor particle search,
which is computationally expensive, we use a mean value
of $(du/dt)_{{\rm ad},i}$ at $t=t_n$ and $t=t_n+\Delta t_{\rm dyn}$.
%\footnote{The code carries out the sub-timestep integration twice
%to update the thermal energy from $t=t_n$ and $t=t_n+\Delta t_{\rm dyn}$.
%First one uses (du/dt)_{{\rm ad},i}^{t_n+\Delta t_{\rm dyn}}(...
Thus, the code re-calculates
only the radiative cooling and heating terms at the sub-time step,
$t+\Delta t_{{\rm sub},i}$, i.e. 
$\Gamma_i^{t+\Delta t_{{\rm sub},i}}$ and 
$\Lambda_i^{t+\Delta t_{{\rm sub},i}}$,
and the other terms are fixed.
In addition, for simplicity we assume that the gas density does not
change dramatically within $\Delta t_{\rm dyn}$; i.e.,
the gas density is assumed to be constant with 
$\rho_{{\rm g},i}^{t_n+\Delta t_{\rm dyn}}$
during these sub-time steps.

\subsection{Yields}

It is well known that the iron yield suggested in \citet{ww95}
seems to be overestimated and leads to much lower [$\alpha$/Fe] values, 
compared to those observed in low-metallicity stars 
in the solar neighborhood. Since the iron yield is the most
ambiguous yield for the SNe II nucleosynthesis model,
it has been discussed that the half of Woosley \& Weaver's suggested yield
is more appropriate \citep[e.g.,][]{tww95}.
Therefore, the updated code adopts half of the yield of
iron in \citet{ww95}.

\subsection{Star formation criteria}

The criteria of star formation are also changed from what
were adopted in \citet{kg03a}. 
In the updated code, the Jeans unstable condition and density threshold 
are excluded from the criteria.
Instead, we introduced a new criterion that
the cooling time has to be smaller than
the dynamical time, i.e.\ $t_{\rm cool} = e/|\Lambda-\Gamma| 
< t_{\rm dyn}=\sqrt{3 \pi /16 G \rho}$ and $\Lambda > \Gamma$.
Finally, this cooling time criterion and 
the convergence of the gas velocity field, 
${\bf \nabla} \cdot \mbox{\boldmath $v$}_i < 0$, are
the criteria for star formation in the updated code. 
The Jeans unstable condition is discarded, because it is sensitive 
to the numerical resolution \citep{ojeqf03}. The density threshold
is ignored, because it is not well understood whether or not
there is such threshold, except for disk galaxies.
In addition,
for disk galaxies the density threshold explained by the dynamical
instability condition of the rotating gas disk \citep{rk89}, which should be
able to be naturally taken into account in dynamical simulations.

\end{document}